\documentclass[prb,aps,twocolumn,amsmath,amssymb,showpacs,nobibnotes]{revtex4}

\usepackage{graphicx}
\usepackage{dcolumn}
\usepackage{bm}

\begin{document}

\title{Effect of oxygen content on the transport properties
and magnetoresistance in [Ca$_{2}$CoO$_{3-\delta}$]$_{0.62}$[CoO$_{2}$]
single crystals}
\author{X. G. Luo, X. H. Chen$^\ast$, G. Y. Wang, C. H. Wang,
Y. M. Xiong, H. B. Song, H. Li and X. X. Lu }
\affiliation{Hefei National Laboratory for Physical Science at
Microscale and Department of Physics, University of Science and
Technology of China, Hefei, Anhui 230026, People's Republic of
China}

\date{\today}

\begin{abstract}
Transport property is investigated in
[Ca$_{2}$CoO$_{3-\delta}$]$_{0.62}$[CoO$_{2}$] single crystals
obtained by varying annealing conditions. The $\rho_{ab}(T)$
exhibits a resistivity minimum, and the temperature corresponding
to this minimum increases with the loss of oxygen content,
indicative of the enhancement of spin density wave (SDW). Large
negative magnetoresistance (MR) was observed in all single
crystals [Ca$_{2}$CoO$_{3-\delta}$]$_{0.62}$[CoO$_{2}$], while a
magnetic-field-driven insulator-to-metal (IM) transition in oxygen
annealed samples. These results suggest a ferromagnetic
correlation in system enhanced by oxygen content. In addition, a
low temperature thermal activation resistivity induced by fields
was observed in single crystals annealed in oxygen atmosphere.
\end{abstract}

\pacs{75.47.-m, 71.30.+h, 75.30.Fv, 75.50.Gg}
 \maketitle

\section{INTRODUCTION}
\vspace*{-2mm}
 Misfit-layered cobalt oxides attracted a great deal of
interest for their possession of unusual electrical properties
such as large negative magnetoresistance,\cite{Masset, Maignan,
Pelloquin,Yamamoto} coherent-incoherent transition with varying
temperature in (Bi,Pb)$_2$Ba$_3$Co$_2$O$_9$,\cite{Valla} and a
large thermoelectric power (TP) with metallic resistivity observed
in (Bi,Pb)$_2$Sr$_2$Co$_2$O$_9$,\cite{Valla, Tsukada}
Ca$_3$Co$_4$O$_9$ ([Ca$_{2}$CoO$_{3-\delta}$]$_{0.62}$[CoO$_{2}$]), \cite{Masset}
and Tl$_{0.4}$(Sr$_{0.9}$O)$_{1.12}$CoO$_2$.\cite{Hebert}
Besides the effort to enhance the thermoelectric figure of merit
ZT=S$^2$T/$\rho\kappa$ for the application reason, as a transition
metal oxide with strong correlation and anomalous electronic
structure, many studies were focused on their physical nature.
Recently, another promising thermoelectric triangular cobaltite
Na$_x$CoO$_2$, was found to exhibit superconductivity below 5 K by
intercalating water molecules into between the Na$^+$ and CoO$_2$ layers
in the composition of $x$=0.35. \cite{Takada} Later, Foo et al.
\cite{Foo} observed an insulating resistivity below 50 K in the
composition of $x$=0.5, which is related to the strong coupling of
the holes and the long-range ordered Na$^+$ ions. The strong
dependence of thermopower on magnetic field in Na$_{x}$CoO$_2$
provides a unambiguous evidence of strong electron-electron
correlation in the thermoelectric cobalt oxides. \cite{Wang} The
large TP with metallic resistivity, superconductivity, charge
ordering existing in various $x$, displays a complicated and
profuse electronic state in Na$_x$CoO$_2$. This has inspired
numerous theoretical and experimental studies on the
thermoelectric cobaltite.

In this paper we focus on one of the thermoelectric cobalt oxides,
[Ca$_{2}$CoO$_{3-\delta}$]$_{0.62}$[CoO$_{2}$] system. Because of
the large ZT ($\sim$1 at 1000 K), \cite{Funahashi}
Ca$_{3}$Co$_{4}$O$_{9}$ is thought to be one of the most promising
candidate for a p-type material component of thermoelectric power
generation systems. [Ca$_{2}$CoO$_{3-\delta}$]$_{0.62}$[CoO$_{2}$]
has a complex magnetic structure, including the spin-state
transition of cobalt ions from low-spin to
intermediate-spin+high-spin or high-spin at about 380 K, the spin
density wave transition with onset temperature $T^{\rm {on}}_{\rm
{SDW}}$=100 K and transition width $\Delta T$=70 K, and the
ferrimagnetism below 19 K.\cite{Masset, Sugiyama} Transport
properties are closely related to the magnetism due to the
coupling between the charge and spin degree of freedom. The spin
state transition of cobalt ions around 380 K leads to an abrupt
jump of resistivity with decreasing temperature; \cite{Masset,
Sugiyama} the emergence of insulator-like behavior below about 70
K was thought to be related to the appearance of short-range SDW
order. \cite{Sugiyama}
[Ca$_{2}$CoO$_{3-\delta}$]$_{0.62}$[CoO$_{2}$] also has a complex
crystal structure. The crystal structure of
[Ca$_{2}$CoO$_{3-\delta}$]$_{0.62}$[CoO$_{2}$] consists of
alternating stacks of two monoclinic subsystems along the c-axis:
the triple Ca$_2$CoO$_{3+\delta}$ rocksalt-type layers and single
CdI$_2$-type CoO$_2$ triangular sheet (conducting
layers).\cite{Masset} The misfit between the two subsystems leads
to an incommensurate (IC) spatial modulation along b axis. The
[Ca$_{2}$CoO$_{3-\delta}$]$_{0.62}$[CoO$_{2}$] system could
exhibit very complicated electronic transport properties due to
the complex magnetic and crystal structure. Strong magnetic field
is a very useful tool to investigate the correlation between
charge dynamics and magnetism. The interesting phenomenon could be
expected under high magnetic field due to the strong coupling
between the charge and spin degrees of freedom. Large negative MR
has been reported in previous work,\cite{Masset} but only on curve
of $\rho(T)$ vs. $T$ under the magnetic field of 7 T was
presented. Detailed investigation is lacking to make out the
mechanism for the MR yet. In this paper, the $\rho_{ab}(T)$ is
studied with varying annealing condition. It is found that the
$\rho_{ab}(T)$ exhibits a minimum. The temperature corresponding
to this minimum ($T_{\rm {min}}$) increases with the loss of
oxygen content, indicative of the enhancement of SDW. The
magnetoresistance were also measured up to 14 T in the
[Ca$_{2}$CoO$_{3-\delta}$]$_{0.62}$[CoO$_{2}$]single crystals.
Large negative MR was observed in all samples. An
insulator-to-metal (IM) transition were observed for oxygen
annealed crystals in high magnetic fields. The
magnetic-field-driven charge dynamics is attributed to the
enhancement of ferromagnetic correlation. \vspace*{-2mm}

\section{EXPERIMENTAL}
\vspace*{-2mm}

The single crystals used in the measurements were grown by the
solution method as described in ref.(13) using K$_2$CO$_3$-KCl as
fluxes with composition of 4:1. The typical dimension of the
crystals are 5$\times$5$\times$0.02 mm$^3$. In order to study the
oxygen content (or carrier concentration) effect on the charge
transport, as-grown crystals were annealed in flowing oxygen
atmosphere at 723 K for 12 h, 11 h, 10 h and 8 h, , which were denoted
as crystal A, B, C and D, respectively. The as-grown crystal is denoted
as crystal E. Some other crystals were annealed in flowing nitrogen
atmosphere at 723 K for 10 h, denoted as crystal F.  We found that
the charge transport is sensitive to the annealing time (oxygen content).
The details will be discussed in the following section. This result is
consistent with the reported by Karppinen et al.\cite{Karppinen}
Resistance measurements were performed using the ac four-probe method
with an ac resistance bridge system (Linear Research, Inc.; LR-700P).
The dc magnetic field for magnetoresistance measurements is supplied
by a superconducting magnet system (Oxford Instruments). In the text
following, we abbreviate [Ca$_{2}$CoO$_{3-\delta}$]$_{0.62}$[CoO$_{2}$]
to Ca$_{3}$Co$_{4}$O$_{9}$ for simplicity.
\vspace*{-2mm}

\section{EXPERIMENTAL RESULTS}
\vspace*{-2mm}

\subsection{Transport properties}
\vspace*{-2mm}

\begin{figure}[h]
\centering
\includegraphics[width=6.5cm]{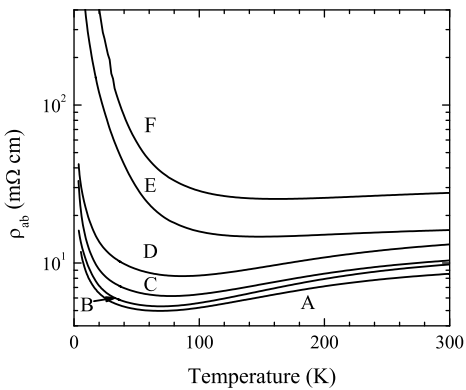}
\vspace*{-2mm}
\caption{The temperature dependence of
resistivity for the Ca$_{3}$Co$_{4}$O$_{9}$ crystals annealed in
various conditions. The curve A, B, C, D correspond to the
resistivity of the crystal annealed in oxygen flow at 723 K for 12
h, 11h, 10h, 8h, respectively. The curve E and F are obtained from
as-grown crystal and nitrogen-annealed crystal, respectively. }
\label{Fig1}
\vspace*{-2mm}
\end{figure}

Figure 1 shows the $\rho(T)$ curves of the Ca$_{3}$Co$_{4}$O$_{9}$
single crystals annealed in various conditions. There are some
common features in $\rho(T)$ among the crystals. The $\rho(T)$ is
metallic in high temperatures and shows a broad minimum with
decreasing temperature. Below the temperature of resistivity
minimum ($T_{\rm {min}}$), the $\rho(T)$ exhibits a diverging behavior.
The crystal A, which was annealed in oxygen flow at 723 K for 12 h has
the lowest resistivity (the curve A) and $T_{\rm {min}}$ (68.8 K) among
the listed samples. The ratio $\rho(T$=4 K)/$\rho(T$=300 K) is also the
smallest. This indicates that this sample corresponds to the highest
hole concentration. With decreasing the
annealing time in oxygen flow at 723 K, $T_{\rm {min}}$ and the ratio
$\rho(T$=4 K)/$\rho(T$=300 K) increases, indicative of the decrease of
the hole concentration. This demonstrates that annealing in oxygen
atmosphere enhances the oxygen content in the crystals. The hole
concentration increases with the enhancement of oxygen content.
The crystal annealed in nitrogen flow exhibits the largest resistivity
among all samples, and it also possesses the largest $T_{\rm {min}}$
($\sim$169 K), indicating that this crystal have the lowest hole
concentrations and oxygen content. The as-grown crystal has the
resistivity and $T_{\rm {min}}$ between that of the oxygen-annealed
crystals and nitrogen-annealed one, manifesting a hole concentration
between them. Thus the annealing in nitrogen flow is a procedure of
reducing oxygen content, while in oxygen flow enhancing oxygen content.
This is consistent with the results reported by
Karppinen et al.\cite{Karppinen} They found
that the oxygen content is easily changed just by ambient pressure in
low oxygen content regime. The $\delta$ difference in
Ca$_{3}$Co$_{4}$O$_{9+\delta}$ is as high as 0.24 between $N_2$
annealed sample and sample prepared in air. However, the annealing
has to be performed at high oxygen pressure to change the oxygen
content in the large $\delta$ regime. The difference between our
and their works is that their work was carried out on the
polycrystalline samples, while the data presented here on single
crystals. Fig. 1 clearly shows that the charge transport is sensitive
to the oxygen content in the crystals, which increases with
enhancing annealing time at 723 K in oxygen flow. In addition, it
can be found that $T_{\rm {min}}$ decreases with increasing the oxygen
content from 149 K in the curve E to 68.8 K in the curve A.
Sugiyama et al. suggested that the resistivity minimum is associated
with the SDW transition. According to this point of view,
the SDW can be enhanced by decreasing the oxygen content,
equivalently, reducing the Co ions valence. This is consistent with the
effect of Bi and Y doping in the Ca$_{3}$Co$_{4}$O$_{9}$, where
the $T^{\rm {on}}_{\rm {SDW}}$ is enhanced by 30 K at the doping
level of 0.3 with decreasing the Co valence. These results indicate
that the SDW is very sensitive to the Co ions valence. It should be
pointed out that annealing in nitrogen may have a much stronger effect
on the enhancement of SDW than the Y and Bi doping because
the $T_{\rm {min}}$ of the curve F is enhanced by about 100 K
relative to that of curve A, though the exact transition temperature
 of SDW cannot be determined.
\vspace*{-2mm}

\subsection{Magnetotransport phenomena}
\vspace*{-2mm}

\subsubsection{Magnetoresistance}\vspace{-2mm}

\begin{figure}[htp]
\centering
\includegraphics[width=6.5cm]{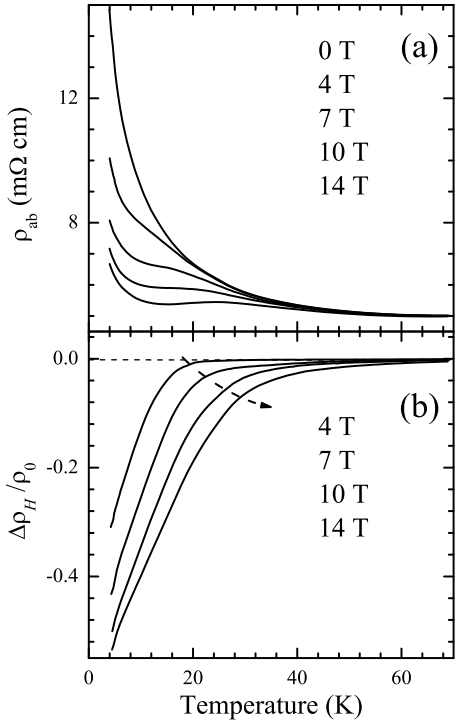}
\vspace*{-2mm} \caption{(a)The temperature dependence of the
resistivity for the crystal A at different fields. (b) The
temperature dependence of $\Delta\rho_H/\rho_0$
=[$\rho(H)-\rho$(0)]/$\rho$(0) at different fields. The dashed
arrow in the (b) guides one's eyes for the temperature below which
the negative MR begins to increase rapidly.} \label{fig2}
\vspace*{-2mm}
\end{figure}

Figure 2(a) shows the in-plane resistivity of the crystal A as the
function of temperature in the range of 4 K to 70 K in various
magnetic fields up to 14 T. This crystal was annealed in oxygen
flow at 723 K for 12 h. The field is applied along the c-axis. At
zero field the sample is metallic (d$\rho_{ab}$/dT$>$0) in high
temperatures, and exhibits a minimum of $\rho(T)$ at 68.8 K. Below
68.8 K the crystal shows an insulator-like behavior. The $\rho(T)$
in whole range of $T$ at zero field has been shown by the curve A
of fig. 1. The value of $\rho_{{ab}}(T)$ is suppressed strongly at
low temperature by magnetic field, with MR
[=($\rho(T,H)$-$\rho(T$,0$)$)/$\rho(T$,0)] reaching $-55\%$ at 4 K
and 14 T. The large negative MR comes from the reduction of spin
scattering, which reflects a spin polarized transport. The
reduction of spin scattering originates from the ferromagnetic
correlation, which can be deduced from the ferrimagnetic
transition at about 19 K. The negative MR shown in fig. 2(b)
increases with decreasing the temperature, in contrast to the CMR
in manganites \cite{Tokura}or ReBaCo$_2$O$_{5+\delta}$,
\cite{Troyanchuk, Kim} in which a maximum of MR can be observed
around the phase transition temperature. This indicates that the
ferromagnetic correlation in the conducting layers is rather weak
or some fluctuation exists in the system. Figure 2(b) shows a
large negative MR below 20 K at 4 T, but no obvious MR above 20 K.
This temperature is almost the same as that of the ferrimagnetic
transition, reflecting the close relation between the spin
polarized transport and the ferromagnetic correlation inferred
from ferrimagnetism. The MR displayed in fig. 2(b) increases
abruptly below a certain temperature at different fields (as
indexed by the dashed arrow), and this temperature increases with
enhancing external magnetic field. Such a temperature should
correspond to the ferrimagnetic transition, which would be
enhanced with increasing magnetic field. These results support
that the large negative MR arises from the ferromagnetic
correlation in conducting layers. \vspace*{-2mm}

\subsubsection{Magnetic-field-induced IM transition}
\vspace*{-2mm}

\begin{figure}[htp]
\centering
\includegraphics[width=6.5cm]{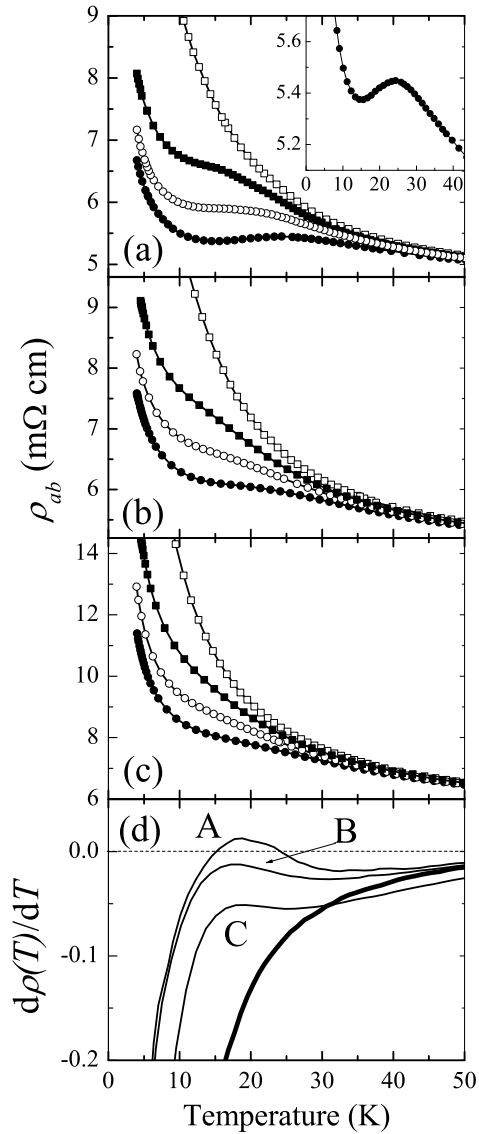}\vspace*{-2mm}
\caption{Magnetoresistivity of the crystal A (a), B (b) and C (c)
plotted against $T$ in detail at magnetic fields of 0 T ({\tiny
$\square$}), 7 T ({\tiny $\blacksquare$}), 10 T ($\circ$) and 14 T
($\bullet$), respectively. (d) The slope of $\rho(T)$ under the
field of 14 T for crystal A, B and C (solid lines) and of
$\rho(T)$ without field (bold line) for crystal A plotted against
$T$.  The inset shows the magnified plot of 14 T data of crystal A
in order to exhibit the IM transition clearly.} \label{fig3}
\vspace*{-2mm}
\end{figure}

A most intriguing result in fig. 2(a) is that a field-induced
weak insulator-to-metal transition at 14 T can be observed at about 25 K.
The data below 50 K at 0, 7, 10, 14 T for crystal A, B, C are plotted
in the fig. 3(a), (b), and (c), respectively. In fig. 3(a), there
exhibits a clear downturn of $\rho(T)$ at about 25 K for 14 T data,
at which the sign of slope d$\rho(T)$/d$T$ is changed
from negative to positive, indicating that an insulator-metal transition is
induced by magnetic field. Compared to the zero field curve, $\rho(T)$
in magnetic fields lower than 14 T shows a notable change of the slope
(d$\rho(T)$/d$T$) at low temperature. Fig. 3(a) shows that even $\rho(T)$
at 4 T begins to possess the marked change of the slope. It is similar to
the cusp structure of $\rho(T)$ of Bi$_{2-x}$Pb$_{x}$Sr$_{2}$Co$_{2}$O$_{y}$
single crystals with x=0.51. In Bi$_{2-x}$Pb$_{x}$Sr$_{2}$Co$_{2}$O$_{y}$
single crystals, the zero field $\rho(T)$ shows an obvious downturn at 5 K
for $x$=0.51, which is thought to correspond to the ferromagnetic
transition.\cite{Yamamoto} The cusp shifts to higher temperature with
increasing magnetic field. This is also similar to the rapidly downturn
of $\rho(T)$ around ferromagnetic transition with decreasing $T$ in the
perovskite manganites\cite{Tokura} and cobalt oxides\cite{Goodenough}.
These suggest that the Ca$_{3}$Co$_{4}$O$_{9}$
system is close to ferromagnetic transition, especially in high magnetic
field, though a ferromagnetic transition has not been found in this system
experimentally yet. In the crystal A, ferrimagnetism instead of
ferromagnetism was observed in weak magnetic field. No measurement
has been performed in high field. Nonetheless, inferred from the a
notable change of the slope of $\rho(T)$ in magnetic field, our results
seem to suggest that ferrimagnetism to ferromagnetism transition possibly
occurs when magnetic field is enhanced.

For the crystal B and C with less oxygen content relative to the
crystal A, no downturn behavior in $\rho(T)$ is observed even at 14 T,
but it shows a remarkable change in the slope d$\rho$/d$T$ at the
ferrimagnetic transition temperature ($T_{\rm FR}$), as shown in fig. 3(b)
and 3(c). In order to show clearly the above behavior, the slope
d$\rho$/d$T$ vs. $T$ curve presented in Fig. 3(d).  The d$\rho(T)$/d$T$
at zero field varies monotonically with temperature, while d$\rho(T)$/d$T$
at 14 T for these three samples show a clear dip-hump structure, which
reveals a noticeable change of the slope of $\rho(T)$ around 10 K-30 K
in magnetic field. The increase of the d$\rho(T)$/d$T$ toward to positive
value with decreasing $T$ indicates a downturn tendency  of $\rho(T)$.
Obviously, the tendency of the downturn of $\rho(T)$ decreases with reducing
the oxygen content. A positive d$\rho$/d$T$ is clearly shown in fig. 3(d)
for the crystal A at 14 T, indicative of an insulator-metal transition.
The field-induced IM transition is ascribed to the enhancement
of ferromagnetic correlation. Thus the ferromagnetic correlation
is reduced with the loss of oxygen content. As shown in fig. 1,
SDW order is enhanced by the reduction of oxygen content. Thus the
ferromagnetic correlation seems to be \emph{competing} with the SDW.

\subsubsection{Magnetotransport in the as-grown crystal}
\vspace*{-2mm}

\begin{figure}[hbtp]
\centering
\includegraphics[width=6.5cm]{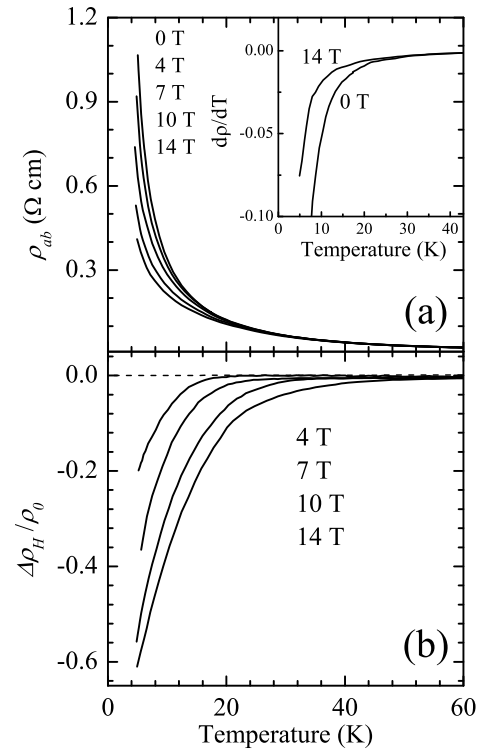}\vspace*{-2mm}
\caption{(a) The temperature dependence of the
transverse magnetoresistivity the in as-grown crystal. (b)
Temperature dependence of the MR in various fields.}
\label{fig4}
\vspace*{-2mm}
\end{figure}

Compared to crystal A-D, the as-grown crystal E has a
lower carrier concentration and thus a higher resistivity.
Figure 4(a) shows the magnetoresistivity for the as-grown crystal
at various magnetic fields. The large negative MR is also observed
at low temperature, like the crystal A-D. While no
sign for a field-driven IM transition can be found even at 14 T
(see the inset of fig. 4(a)). A reduction of carrier
concentration in the as-grown crystal relative to the oxygen annealed
ones could lead to an enhancement of SDW order as in the Bi- or
Y-for-Ca substitution in Ca$_{3}$Co$_{4}$O$_{9+\delta}$
samples.\cite{Sugiyama} In the literature, it is reported that
the SDW transition temperature is increased by $\sim$ 30 K due to
either Bi or Y doping with $x$=0.3 in
Ca$_{3-x}$M$_{x}$Co$_{4}$O$_{9}$ (M=Bi,Y). The $T_{\rm {min}}$ for the
as-grown crystal at zero field is 149 K, which is much higher
than that observed in the crystals annealed in oxygen flow. T$_{\rm {min}}$
is associated to the SDW order.\cite{Sugiyama} Thus the increase of
$T_{\rm {min}}$ is consistent with the enhancement of the SDW due to
reduction of carrier concentration. In fig. 4(b), the
$\Delta\rho _{H}/\rho _{0}$ shows negative value, indicative of a spin
polarized transport like in crystal A. One can notice that in fig. 4(b)
the temperature for the $\Delta\rho_{H}/\rho_{0}$ beginning to show
negative value enhances with increasing magnetic field, suggesting a
enhancement of ferromagnetic correlation with increasing magnetic field
as in crystal A. Nonetheless, the ferromagnetic correlation is much
weaker than that in the crystals annealed in oxygen so that no sign of
downturn tendency of $\rho(T)$ can be found at 14 T
in spite of the large negative MR. While a larger negative MR relative to
crystal A is observed at 14 T in crystal E. The possible reason for this
result will be discussed in the following.

\subsubsection{The effect of annealing condition on the transport
behavior}\vspace*{-2mm}

\begin{figure}[htp]
\centering
\includegraphics[width=6.5cm]{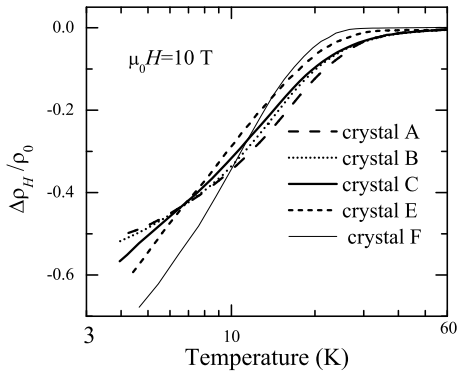}
\vspace*{-2mm}
\caption{The magnetoresistance at 10 T as a function of
\emph{T} in the crystal A-C, E, and F.}
\label{fig5}
\vspace*{-2mm}
\end{figure}

According to the results above, the annealing condition affects the
transport and magnetotransport behaviors strongly. Obviously,
the transport behavior is changed by varying oxygen content, i.e.
by varying the Co ions valence. From crystal A to F, the evolution
of resistivity indicates that the hole concentration is reduced
due to the loss of oxygen content. The reduction of the Co ions
valence due to the loss of oxygen content enhances the SDW order
dramatically, which can be inferred from the large increase of
the $T_{\rm {min}}$ as discussed above. Spin density wave is
basically antiferromagnetic, thus one can conclude that there is
antiferromagnetic correlation in the system and the loss of the oxygen
favors the enhancement of the antiferromagnetic correlation. Nonetheless,
the ferrimagnetism and the large negative MR in the crystals suggest
that there also exists ferromagnetic correlation.
Figure 5 exhibits the $\Delta\rho(H=10$T)/$\rho(H=0$T$)$
vs. $T$ for crystals A-C, E, and F, respectively. Clearly, the
temperature below which the negative MR is observable increases with
enhancing oxygen content. At very low temperature, the absolute value
of $\Delta\rho_{H}/\rho_{0}$ increases with reducing
the oxygen content. With increasing $T$, it gradually evolves to a
completely opposite variation with oxygen content, that is, the absolute
value of $\Delta\rho_{H}/\rho_{0}$ decreases with reducting
the oxygen content. This is very similar to the evolution of
 $\Delta\rho_{H}/\rho_{0}$ with lead doping level in
(Bi,Pb)$_{2}$Sr$_{2}$Co$_{2}$O$_{y}$ system (Ref. 4, fig. 9).
With increasing Pb doping level, the temperature below which negative MR
is observable increases, and the value of MR is enhanced in the temperature
range from 8 to 40 K, while reduced below 8 K.\cite{Yamamoto}
In (Bi,Pb)$_{2}$Sr$_{2}$Co$_{2}$O$_{y}$, the increase of hole concentration
due to Pb doping leads to a transition from paramagnetism in Pb-free sample
to weak ferromagnetism in the sample with Pb=0.51.\cite{Yamamoto}
Thus the fig. 5 suggests that the enhancement of oxygen content, i.e.
the increase of Co ions valence, leads to the increase of the
ferromagnetic spin correlation. This is consistent with conclusion
inferred from the evolution of the field-induced IM transition by
comparing the fig. 3 and fig. 4.

\begin{figure}[htp]
\centering
\includegraphics[width=0.45\textwidth]{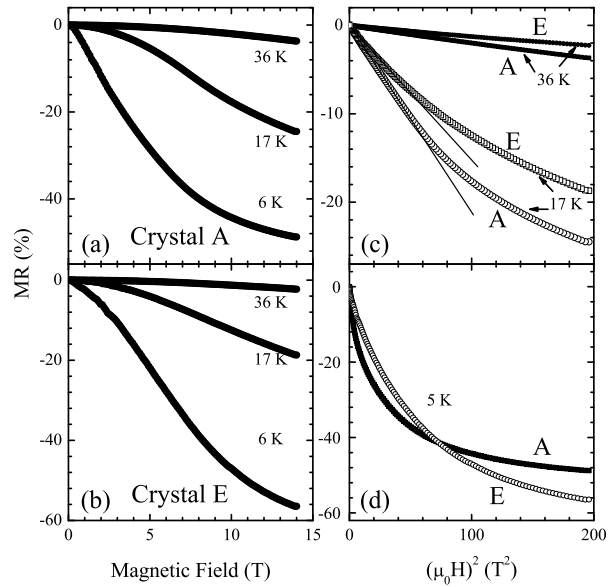}
\caption{The isothermal magnetoresistance as a function of
magnetic field for crystal A (a) and E (b), respectively. The same data are
replotted in (c) and (d) in the square magnetic field scale.}
\label{fig6}
\vspace*{-2mm}
\end{figure}

The isothermal MR shown in fig. 6(a) and 6(b) further confirms this
speculation. Fig. 6(a) and 6(b) shows the isothermal MR for the
crystal A and E, respectively.  At 36 K and 17 K, the MR is larger
in crystal A. While at 6 K, the MR in crystal A is smaller
than that in crystal E below about 8.5 T, and with further increasing
magnetic field, the MR in crystal A becomes the smaller one (also
see fig. 6(d)). At 36 K and 17 K, the MR shows no saturation tendency,
indicating that the spins is far from being completely polarized.
The larger MR value in crystal A at 36 K and 17 K suggests a more
polarized transport and thus a stronger ferromagnetic correlation.
At 6 K, there is a clear saturation tendency of MR in high magnetic field
in both crystals. The larger MR in lower field in crystal A indicates
that the spin in this sample is easier to be polarized, since the
magnetotransport is thought to be a spin polarized transport.
The saturation of MR suggests a tendency of completely polarized spins.
The smaller MR in crystal A at 6 K in high magnetic
field suggests a less spin disordered transport. All of these strongly
suggest a stronger ferromagnetic correlation in crystal A and definitely
support the speculation that the enhancement of oxygen content leads to
the increase of the ferromagnetic spin correlation.

Fig. 6(a) and 6(b) is replotted in fig. 6(c) and 6(d) in square
magnetic field scale. At 36 K, it is found that the MR of the two
crystals is proportional to $H^{2}$. While the MR at 17 K is
proportional to $H^{2}$ only in low magnetic fields and deviates
from $H^{2}$ law in high magnetic fields. At the lower $T$, 6 K,
it disobeys $H^{2}$ law from zero field. These results suggest
that there are probably different mechanisms for the negative MR
at high $T$ and low $T$.  It should be pointed out that at 36 K
there is (short-range) IC-SDW order only, while 6 K is far below
the ferrimagnetic transition, at which the the ferrimagnetism and
(long-range) IC-SDW order coexist. consequently, the MR at 36 K
only can reflect the property related to SDW state. While the MR
at 6 K should mostly reveal the effect of ferrimagnetism. 17 K is
just below the ferrimagnetic transition, thus the MR at this
temperature involves the influence of these two magnetisms.
\vspace*{-2mm}

\subsubsection{The influence of oxygen content and magnetic field on
the SDW order}\vspace*{-2mm}

\begin{figure}[htp]
\centering
\includegraphics[width=6.5cm]{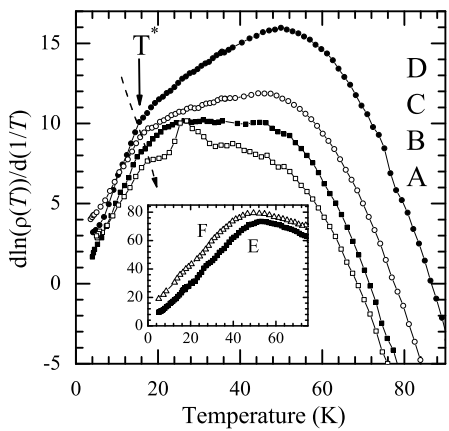}\vspace{-2mm}
\caption{Logarithmic derivatives of the
zero-field resistivity vs. T for crystal A, B, C, and D respectively.
The inset shows the $T$ dependence of the Logarithmic derivatives of the
zero-field resistivity of crystal E.}
\label{fig7}
\vspace{-2mm}
\end{figure}

Why negative MR can be observed in IC-SDW  state? In previous SDW
materials, such as organic conductors (TMTSF)$_2$X (TMTSF:
Tetrmethltetraselena-fulvalene, X=PF$_6$, NO$_3$ and ClO$_4$),
\cite{Matsunaga,Nomura,Biskup} were reported to show positive MR in SDW
state. In order to investigate the nature of such negative MR, the effect
of a magnetic field on the SDW order should be performed. Usually, in a SDW
state, the resistivity $\rho(T)$ can be expressed as \cite{Sugiyama,Gruner}
\begin{equation}
\frac{1}{\rho}=\mu(T)exp\left(-\frac{\Delta_g}{k_BT} \right),
\end{equation}
where $\mu(T)$ is the mobility of carriers, $\Delta_g$ is the gap
energy, and $k_B$ is the Boltzmann constant. Usually, $\mu(T)$ is
independent of $T$. Using Eq. (1), the $\rho(T)$ of crystals A-D
in $T$ between 50 and 20-30 K can be fitted by assuming that
$\mu(T)$ is independent of $T$. In this case, dln$\rho$/d($1/T$)
can be calculated to reflect the activation energy,
$\Delta_g$.\cite{Okabe} Fig. 7 shows the dln$\rho$/d(1/$T$)
plotted against $T$ for all the crystals. Plateaus are observed in
crystals A-D in the $T$ ranging from 20-50 K. This reveals that in
these samples there are $T$ independent gap within this
temperature range, which could be opened by SDW. Fig. 7 exhibits
that the value of the plateau of dln$\rho/$d(1/$T$) increases from
crystal A to D. This is consistent with the evolution of $T_{\rm
min}$ with the oxygen content, which reflects the SDW is enhanced
by decreasing oxygen content. Thus it may suggest that the
observed energy gap is opened by SDW.\cite{Gruner} According to
$\mu$SR and $\mu^{+}$SR results,\cite{Sugiyama} it is considered
that a short-range order IC-SDW state appears below 100 K, while
the long-range order is completed below 30 K. In fig. 7,
dln$\rho$/d(1/$T$) varies smoothly through 30 K with decreasing
temperature, except for the crystal A. In crystal A, a sharp peak
is observed around 27 K. The susceptibility and $\mu^{+}$SR
experiment reveal that the long-range SDW order completes at this
temperature.\cite{Sugiyama} Moreover, the sharp peak of
dln$\rho$/d(1/$T$) was thought to correspond to the SDW transition
temperature.\cite{Matsunaga,Nomura,Biskup} Thus, the peak of
dln$\rho$/d(1/$T$) in crystal A corresponds to the temperature
where long-range SDW order completes. Even so, the value of
dln$\rho$/d(1/$T$) is almost the same at the two sides of the
peak. Fig. 7 indicates that the influence of short-range and
long-range IC-SDW order in this sample on the transport properties
is the same.

The Eq. (1) cannot be applied in the whole range of $T$ below 50 K
because the dln$\rho$/d(1/$T$) shows a rapidly decrease below a
temperature around 20 K, which is indexed as $T^\ast$ in fig. 7.
According to the susceptibility experiment, \cite{Sugiyama} a
ferrimagnetic transition occurs around this temperature. This
suggests that the abrupt decrease of dln$\rho$/d(1/$T$) below
$T^\ast$ arises from the ferrimagnetic transition. With increasing
oxygen content from crystal D to crystal A, the $T^\ast$ enhances
slightly. The dln$\rho$/d(1/$T$) reflects the activation gap.
Thus, the abrupt drop of dln$\rho$/d(1/$T$) suggests that the
ferrimagnetic transition reduces or hides the SDW gap. The inset
shows the dln$\rho$/d(1/$T$) of crystal E and F. In this two
samples no plateau is observed, suggesting Eq. (1) is not feasible
any more or the $\mu$($T$) is dependent on $T$ in this two
samples. There is no anomaly below 20 K, which could be due to the
stronger SDW order and rather weaker ferrimagnetism in these two
samples relative to the crystals annealed in oxygen atmosphere.
The effect of ferrimagnetism is hidden by the strong SDW order.

\begin{figure}[htp]
\centering
\includegraphics[width=6.5cm]{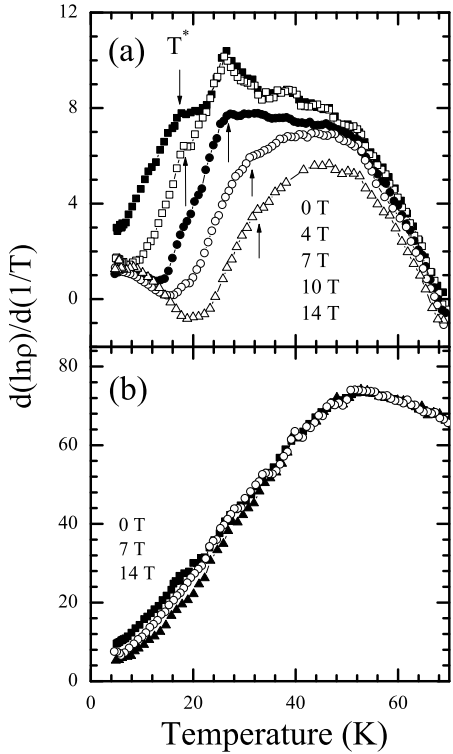}\vspace{-2mm}
\caption{(a) The logarithmic derivative of the
resistivity of crystal A as a function of the temperature at 0, 4, 7, 10,
and 14 T. (b) The temperature dependence of The logarithmic derivative of
the resistivity of crystal E at 0, 7, 14 T.}
\label{fig8}
\vspace{-2mm}
\end{figure}

In order to study the influence of magnetic field on the SDW
state, the dln$\rho$/d(1/$T$) as a function of $T$ at various
magnetic field in the two samples is shown in Fig. 8(a). It is
observed that the value of the plateau decreases with increasing
field, indicating the reduction of the activation gap. While above
50 K, the dln$\rho$/d(1/$T$) is almost independent of magnetic
field. As discussed above, the gap opening can be ascribed to the
SDW order. If so, it suggests that the SDW order here is
suppressed by magnetic field. The peak around 27 K disappears when
field is as high as 7 T, suggesting that the SDW is suppressed
indeed. This can be used to explain the negative MR far above the
ferrimagnetic transition temperature, which has a different $H$
dependence of the MR from that in ferrimagnetic state. Since the
upturn resistivity below $T_{\bf min}$ is associated with the SDW
order, a suppression of SDW certainly leads to negative MR.  In
fig. 8(a), $T^\ast$ increases with increasing magnetic field,
while the slope of the dln$\rho$/d(1/$T$) remains unchanged below
$T^\ast$. The enhancement of the $T^\ast$ is related to the
increase of ferrimagnetic transition temperature in magnetic
field, consistent with the MR data in Fig. 2(b). Clearly,
\emph{the ferrimagnetic ground state is competing with the SDW
ground state}. The suppression of SDW in high magnetic field may
be related to the enhancement of the ferromagnetic correlation.
Another triangular layered cobaltite should be mentioned to
compare with [Ca$_{2}$CoO$_{3}$]$_{0.62}$[CoO$_{2}$] is
Na$_{x}$CoO$_{2}$. In Na$_{0.75}$CoO$_{2}$, SDW and weakly
ferromagnetism coexist,\cite{Sugiyama1,Boothroyd,Sales} while a
large positive MR was observed in magnetic state,\cite{Motohashi}
contrasting to the negative MR in
[Ca$_{2}$CoO$_{3}$]$_{0.62}$[CoO$_{2}$]. The negative MR in
[Ca$_{2}$CoO$_{3}$]$_{0.62}$[CoO$_{2}$] originates from the
reduction of the spin scattering of carriers, while in
Na$_{0.75}$CoO$_{2}$, the MR seems to be more possible coming from
the change of carrier-mobility in Fermi surfaces in magnetic field
due to a pseudogap formation.\cite{Motohashi} In spite of the
weakly ferromagnetism, the pseudogap formation leads to a positive
MR in the Na$_{0.75}$CoO$_{2}$. Finally, we note that the
dln$\rho$/d(1/$T$) vs. $T$ curve in fig. 8(b) for crystal E
remains unchanged above 20 K with varying magnetic field,
suggesting that the SDW is too strong to be suppressed by magnetic
field. Below 20 K, the dln$\rho$/d(1/$T$) decreases with
increasing magnetic field, manifesting that there still is
ferrimagnetic transition in crystal E in spite of the reduction of
oxygen content with respect to the crystal A. This further
demonstrates the point of veiw that the ferromagnetic correlation
leads to the large negative MR.\vspace*{-2mm}

\subsection{Thermal-activation transport in magnetic field at low
temperature}
\vspace*{-2mm}

\begin{figure}[htp]
\centering
\includegraphics[width=6.5cm]{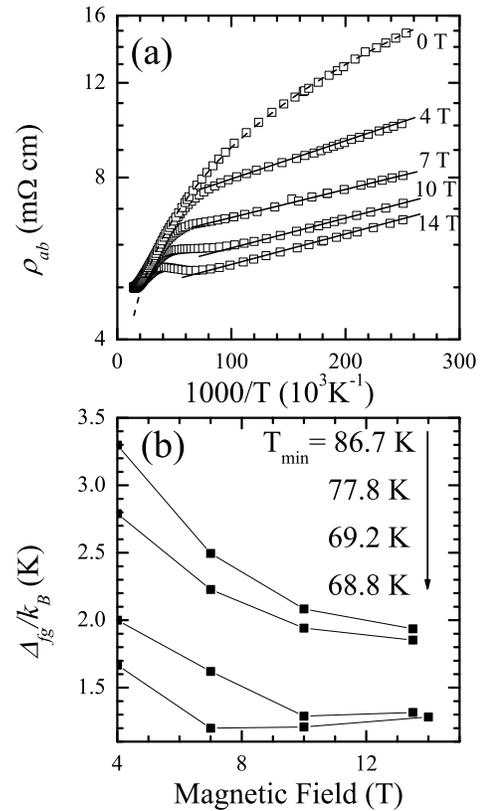}\vspace*{-2mm}
\caption{ (a) Arrhenius plot for the data of fig.
1(a). The dashed line is the fitting data with Mott variable range
hopping law. The solid line is the fit data with the thermal
activation formalism. (b) The field-induced energy gap
($\Delta_{fg}$) plotted against magnetic field for four crystals
with different T$_{\rm {min}}$.}
\label{fig9}
\vspace*{-2mm}
\end{figure}

In Fig. 8(a) another intriguing feature can be found in the low
temperature range, the dln$\rho$/d($1/T)$ shows a $T$-independent
behavior below about 10 K for the data in magnetic field. It
suggests that there is another thermal activation transport. The
Arrhenius plot of the data in fig. 1(a) is displayed in fig. 9(a),
which clearly shows that the temperature dependence of the
resistivity under different fields exhibits an Arrhenius-type
behavior below about 12 K, that is, $\rho(T)\propto
e^{\Delta/k_BT}$, where $\Delta$ is gap energy and $k_B$ is
Boltzmann constant. It suggests that the resistivity in low
temperature range under the fields follows the thermal activation
model. While the zero field data below 30 K can be well fitted by
a formula $\rho$(T)=Ae$^{(T_0/T)^{1/4}}$, indicative of Mott
variable range hopping (VRH) law. It implies that an energy gap is
opened by magnetic field. It should be pointed out that this
phenomenon was observed in all of the four oxygen annealed
crystals A-D, with different $T_{\rm {min}}$. The obtained gap
energy depends on $T_{\rm {min}}$. Figure 9(b) shows the magnetic
dependence of the field-induced gap energy for the four crystals.
It is found that the gap energy increases with increasing $T_{\rm
{min}}$. While it decreases with increasing magnetic fields, and
saturates in high field. It should be pointed out that no such
phenomenon is observed in the as-grown crystal and the sample
annealed in $N_2$. The nature of this field-induced thermal
activation transport is not well understood yet. The gap energy
shown in fig. 9(b) is dependent on $T_{\rm {min}}$, suggesting
that the gap may be associated with the SDW order.

\vspace*{-2mm}

\section{DISCUSSION}
\vspace*{-2mm}

Large negative MR has been observed in many misfit-layered
cobaltites, such as: $(Bi,Pb)_2Sr_2Co_2O_y$,
$[Sr_{1.9}Pb_{0.7}Co_{0.4}O_3][CoO_2]_{1.8}$,
$[Bi_{1.7}Co_{0.3}Ca_2O_4][CoO_2]_{1.67}$, and
$[Ca_2CoO_{3-\delta}]_{0.62}[CoO_{2}]$ studied in this paper. In
some of these cobaltites, such as
$[Sr_{1.9}Pb_{0.7}Co_{0.4}O_3][CoO_2]_{1.8}$ and
$[Bi_{1.7}Co_{0.3}Ca_2O_4][CoO_2]_{1.67}$, spin density wave was
observed.\cite{Sugiyama1} While in some others, like
(Bi,Pb)$_{2}$Sr$_{2}$Co$_{2}$O$_{y}$, ferromagnetic transition was
observed. In [Ca$_{2}$CoO$_{3-\delta}$]$_{0.62}$[CoO$_{2}$] single
crystals, spin density wave and ferrimagnetic transition were
observed by $\mu^{+}$SR and susceptibility
experiments,\cite{Sugiyama} respectively.

 Na$_{x}$CoO$_{2}$ is the most
extensively studied at experimental and theoretical level in the
layered cobaltites with triangular [CoO$_{2}$] layers as basic
structure units. Because of the very similar structure of
[CoO$_{2}$] layers, most of the results obtained in
Na$_{x}$CoO$_{2}$ can be reproduced in the misfit cobaltites. The
low-spin state of Co$^{3+}$ and Co$^{4+}$ predicted by Singh in
Na$_{x}$CoO$_{2}$,\cite{Singh} is found to hold in
(Bi,Pb)$_{2}$Sr$_{2}$Co$_{2}$O$_{y}$ by photoemission
experiment.\cite{Mizokawa} Another prediction by Singh in
Na$_{x}$CoO$_{2}$, that the $t_{2g}$ orbits are split into
$a_{1g}$ and $e'_{g}$ subbands due to rhombohedral crystal
field,\cite{Singh} is feasible in
(Bi,Pb)$_{2}$Sr$_{2}$Co$_{2}$O$_{y}$.
 Obviously, some theoretical results obtained in Na$_{x}$CoO$_{2}$ can be
 used in the misfit cobaltites.

A very important prediction from local spin density approximation
(LSDA) band calculation by Singh is the ferromagnetic instability
in Na$_{x}$CoO$_{2}$.\cite{Singh, Singh1} While experimentally,
ferromagnetism has not been observed. Only weakly ferromagnetism
\cite{Sugiyama1,Sales,Boothroyd} or metamagnetism\cite{Luo} was
observed, which is probably due to quantum
fluctuation.\cite{Boothroyd,Singh,Singh1} Neutron inelastic
scattering gives evidences for the existence of ferromagnetic (FM)
spin fluctuation within the [CoO$_{2}$] layers in
Na$_{0.75}$CoO$_{2}$ single crystal.\cite{Boothroyd} The weak
ferromagnetism in Na$_{0.75}$CoO$_{2}$ was thought to originate
from a canted-SDW \cite{Sugiyama1,Sales} or a spin arrangement
with an in-plane ferromagnetic order and a SDW modulation
perpendicular to the planes.\cite{Boothroyd} Metamagnetism in
Na$_{0.85}$CoO$_{2}$ was suggested to favor the in-plane
ferromagnetic correlation and interplane antiferromagnetic (AF)
coupling.\cite{Luo} The Metamagnetism was thought to be a
spin-flop transition from an AF to a FM state along
c-axis.\cite{Luo} The ferromagnetic spin correlation is very
sensitive to the Co ions valence in the triangular cobaltites. In
(Bi,Pb)$_{2}$Sr$_{2}$Co$_{2}$O$_{y}$,\cite{Yamamoto} with
increasing Pb doping level from zero to 0.51, it can change from
paramagnetism to  weak ferromagnetism. Though no ferromagnetism is
observed in [Ca$_{2}$CoO$_{3-\delta}$]$_{0.62}$[CoO$_{2}$], it
shows similar MR behavior to (Bi,Pb)$_{2}$Sr$_{2}$Co$_{2}$O$_{y}$,
especially the variation of the MR with oxygen content in fig. 5
is very similar to the evolution of MR with Pb doping level in
(Bi,Pb)$_{2}$Sr$_{2}$Co$_{2}$O$_{y}$. In the
[Ca$_{2}$CoO$_{3-\delta}$]$_{0.62}$[CoO$_{2}$] crystals, a
ferrimagnetism was found below 19 K in susceptibility experiment,
suggesting the ferromagnetic correlation in conducting $CoO_2$
layers safely. The evolution of MR with oxygen content and the
field-induced downturn of $\rho(T)$ indicate that the
ferromagnetic correlation varies dramatically with Co ions
valence. Especially, the field-induced IM transition in crystal A
occurs at the ferrimagnetic transition. This is similar to the
insulator-metal transition induced by field at the ferromagnetic
transition in (Bi,Pb)$_{2}$Sr$_{2}$Co$_{2}$O$_{y}$ with
Pb=0.51.\cite{Yamamoto}  The Na$_{0.75}$CoO$_{2}$, which possesses
weak ferromagnetism,\cite{Sugiyama1,Sales,Boothroyd,Motohashi,Luo}
exhibits large positive instead of negative MR observed in
(Bi,Pb)$_{2}$Sr$_{2}$Co$_{2}$O$_{y}$. This was considered to be
due to a pseudogap formation deduced from the abrupt downturn of
resistivity.\cite{Motohashi}

Another interesting magnetic property in this system is spin
density wave. The SDW emerges from much higher temperature than
ferrimagnetism. \cite{Sugiyama} According to our data, the
negative MR appears far below $T_{\rm min}$, and it coincides with
the ferrimagnetic transition. In contrast to ferromagnetic
correlation, the SDW has a complete opposite response to Co ion
valence. The SDW enhances with reducing Co ion valence according
to the change of $T_{\rm min}$. This indicates that the
ferrimagnetism and the SDW order are \emph{competing}.

It is worthy of note that the SDW and ferromagnetic correlation
affect the transport and magnetotransport properties strongly,
suggesting that the SDW and ferromagnetic correlation coexist in
the [CoO$_{2}$] conducting sheets below ferrimagnetic transition
temperature. The $\mu^{+}$SR experiments have suggested the
coexistence of the IC-SDW state and ferrimagnetism below 19
K.\cite{Sugiyama} Sugiyama et al. suggested that the two
subsystems of the crystal structure (the [Ca$_2$CoO$_{3-\delta}$]
rocksalt-type layers and the [CoO$_{2}$] triangular sheets) act
directly as the two magnetic sublattices for the
ferrimagnetism.\cite{Sugiyama} From this picture, one can
naturally infer the existence of ferromagnetic spin arrangement in
the [CoO$_{2}$] sheets. Large negative MR and the field-induced IM
transition seems to support this picture for the ferromagnetic
correlation in the conducting [CoO$_{2}$] sheets. However, we
cannot give detailed picture how the spin arrangement for the
coexistence of ferrimagnetism and the (long-range order) IC-SDW
below $T_{\rm FR}$ because of the absence of the microscopic
information of the spins in Co ions. Nonetheless, in
[Ca$_{2}$CoO$_{3-\delta}$]$_{0.62}$[CoO$_{2}$], it cannot be a
canted-SDW as that found in
Na$_{0.75}$CoO$_{2}$\cite{Sugiyama1,Sales} because the spin
moments of both SDW and ferrimagnetism are along c-axis and cannot
be detected in the [CoO$_{2}$] planes according to $\mu^{+}$SR and
susceptibility experiment. Also it is very difficult to expect the
[Ca$_{2}$CoO$_{3-\delta}$]$_{0.62}$[CoO$_{2}$] with in-plane
ferroamgnetic order and a SDW modulation perpendicular to the
planes as proposed by another authors in
Na$_{0.75}$CoO$_{2}$\cite{Boothroyd} because of the misfit
structure between [Ca$_{2}$CoO$_{3-\delta}$] and [CoO$_{2}$]
subsystems. Another possible picture is that the SDW order is
ferromagnetic, while the SDW order is basically antiferromagnetic.
Thus, a macroscopic magnetism should be ferrimagnetic, as actually
observed below 19 K. Co-NMR and neutron-scattering experiments are
desired to make out the exact magnetic structure of
[Ca$_{2}$CoO$_{3-\delta}$]$_{0.62}$[CoO$_{2}$].

\vspace*{-2mm}

\section{CONCLUSION}
\vspace*{-2mm}

To summarize, the in-plane resistivity was measured in
[Ca$_{2}$CoO$_{3-\delta}$]$_{0.62}$[CoO$_{2}$] single crystals in
magnetic fields up to 14 T. The zero field $\rho_{ab}(T)$ exhibits
a minimum and $T_{\rm min}$ increases with decreasing the oxygen
content, indicative of the enhancement of SDW. The SDW strongly
depends on the carrier concentration or oxygen content. Large
negative MR is observed for all crystals. While a field-induced IM
transition in high magnetic field were observed only in the
crystal annealed in oxygen atmosphere. These results suggests
there is ferromagnetic correlation in the system and the
correlation is enhanced by increasing oxygen content and magnetic
field . The strong effect of oxygen content on the transport and
magnetotransport behavior implies that both the SDW and
ferrimagnetism locate within the conducting [CoO$_2$] sheets.
\vspace*{-2mm}

\section{ACKNOWLEDGEMENT}
\vspace*{-2mm}

This work is supported by the grant from the Nature Science Foundation of
China and by the Ministry of Science and
Technology of China (Grant No. NKBRSF-G1999064601), the Knowledge
Innovation Project of Chinese Academy of Sciences.\\

\vspace*{5mm} $^{\ast}$ Electronic address: chenxh@ustc.edu.cn

\end{document}